\documentclass[12pt,aps,preprint,nofootinbib,superscriptaddress,nobalancelastpage]{revtex4}
\usepackage{amsmath,amssymb}
\usepackage{graphicx}
\usepackage{multirow}
\usepackage{hyperref}
\usepackage{epsfig}
\usepackage{amsfonts}
\usepackage{bbm}
\usepackage{mathrsfs}
\usepackage{color}

\newcommand{\be}{\begin{equation}}
\newcommand{\ee}{\end{equation}}
\newcommand{\bea}{\begin{eqnarray}}
\newcommand{\eea}{\end{eqnarray}}

\newcommand{\ba}{\begin{array}}
\newcommand{\ea}{\end{array}}
\newcommand{\balg}{\begin{align}}
\newcommand{\ealg}{\end{align}}

\newcommand{\fig}{Fig.}

\newcommand{\tab}{Tab.}
\newcommand{\BR}{{\rm BR}}

\newcommand{\kton}{kt}
\newcommand{\br}[1]{\sigma{\rm Br}_{#1}}

\newcommand{\DM}{{\rm DM}}
\newcommand{\DMbar}{\overline{\rm DM}}

\newcommand{\lsim}
{\raise0.3ex\hbox{$\;<$\kern-0.75em\raise-1.1ex\hbox{$\sim\;$}}}
\newcommand{\gsim}
{\raise0.3ex\hbox{$\;>$\kern-0.75em\raise-1.1ex\hbox{$\sim\;$}}}

\begin{document} 

\title{Probing the Dark Matter mass and nature with neutrinos}

\author{Mattias Blennow}
\email[]{emb@kth.se}
\affiliation{Department of Theoretical Physics,\\ School of Engineering Sciences, KTH Royal Institute of Technology \\
AlbaNova University Center, 106 91 Stockholm, Sweden}

\author{Marcus Carrigan}
\email[]{carri@kth.se}
\affiliation{Department of Theoretical Physics,\\ School of Engineering Sciences, KTH Royal Institute of Technology \\
AlbaNova University Center, 106 91 Stockholm, Sweden}

\author{Enrique Fernandez Martinez}
\email[]{enrique.fernandez-martinez@uam.es}
\affiliation{Departamento and Instituto de F\'{\i}sica Te\'orica UAM/CSIC,\\ Calle Nicol\'as Cabrera 13-15, Cantoblanco, E-28049 Madrid, Spain \\}

\begin{abstract}
We study the possible indirect neutrino signal from dark matter annihilations inside the solar interior for relatively light dark matter masses in the $\mathcal O(10)$~GeV range. Due to their excellent energy reconstruction capabilities, we focus on the detection of this flux in liquid argon or magnetized iron calorimeter detectors, proposed for the next generation of far detectors of neutrino oscillation experiments and neutrino telescopes. The aim of the study is to probe the ability of these detectors to determine fundamental properties of the dark matter nature such as its mass or its relative annihilation branching fractions to different channels. We find that these detectors will be able to accurately measure the dark matter mass as long as the dark matter annihilations have a significant branching into the neutrino or at least the $\tau$ channel. We have also discovered degeneracies between different dark matter masses and annihilation channels, where a hard $\tau$ channel spectrum for a lower dark matter mass may mimic that of a softer quark channel spectrum for a larger dark matter mass. Finally, we discuss the sensitivity of the detectors to the different branching ratios and find that it is between one and two orders of magnitude better than the current bounds from those coming from analysis of Super-Kamiokande data.
\end{abstract}


\preprint{FTUAM-13-2; IFT-UAM/CSIC-13-027}

\maketitle

%%%%%%%%%%%%%%%%%%%%%%%%%%%%%%%%%%%%%%%%%%%%%%%%%%%%%%%%%%%%
\section{Introduction}
\label{sec:introduction}
%%%%%%%%%%%%%%%%%%%%%%%%%%%%%%%%%%%%%%%%%%%%%%%%%%%%%%%%

The first exciting years of data from Large Hadron Collider (LHC) have provided us with the discovery of a Higgs boson that so far fits very accurately the properties expected from the Standard Model (SM)~\cite{Aad:2012tfa,Chatrchyan:2012ufa}. Unfortunatelly, no significant deviation signaling physics beyond the SM has been found yet. Thus, the existence of dark matter (DM) remains one of our few confirmed signals of new physics. Sadly, all our present evidence for DM stems from its gravitational effects and thus we remain ignorant of its fundamental particle physics properties, such as its mass or couplings to other particles, that could help us understand the nature of DM. For this task, the LHC is still a very promising tool, as testified by the significant regions of the possible DM parameter space that it has already probed. In this work we will instead study a very complementary probe to the LHC searches which also has the potential to discriminate the DM mass and interactions: its indirect detection via annihilations in the solar interior giving rise to energetic neutrino fluxes~\cite{Silk:1985ax,Krauss:1985ks,Gaisser:1986ha}.   

For weakly interacting massive particles in the mass range between 0.1--1~TeV, such neutrinos could be detected by neutrino telescopes~\cite{Boliev:1995xz,Barger:2001ur,Ackermann:2005fr,Halzen:2005ar,Abbasi:2009uz}, such as IceCube. In addition, there have been several works~\cite{Bueno:2004dv,Mena:2007ty,Kumar:2011hi,Agarwalla:2011yy,Kumar:2012uh,Boliev:2013ai} on how well the same study would work for lighter DM with a mass of $\mathcal O(10)$~GeV, and some bounds have been obtained for the Super-Kamiokande detector~\cite{Desai:2004pq,Niro:2009mw,Kappl:2011kz}. These studies have focused on the prospects of running or planned detectors typically considered as far detectors of long baseline experiments. The detector technologies under consideration have been ranging from water \v{C}erenkov detectors to liquid scintillator, magnetized iron (MIND) and liquid argon (LAr).

A relatively common trait of the studies performed thus far is the assumption that, since the neutrino energy spectrum has a sharp cutoff at the DM mass, the DM mass will be very well determined by any experiment which is sensitive enough to detect the signal. However, two facts can hinder the DM mass reconstruction: (i) The neutrino energy will not be perfectly reconstructed by the experiment. (ii) If the neutrinos are not produced through direct annihilation into neutrinos, but rather through the decay products of other annihilation channels, then the neutrino spectrum will be softer and it will not always be easy to observe the cutoff energy. This is particularly true when there is a significant amount of background. The determination of the DM mass at the Deep Core experiment has been studied separately for the $\tau$ and $b$ channels~\cite{Das:2011yr}. The aim of the present study is to study the simultaneous determination of the DM branching ratios and the DM mass at future neutrino detectors.

The remainder of this work is organized as follows: In Sec.~\ref{sec:background}, we briefly review the physics of indirect neutrino signals from the Sun, starting from the gravitational trapping of DM inside the Sun, via the generation of neutrino events, to the neutrino propagation to and interaction in a detector on Earth. We continue by discussing the detector technologies used in the study in Sec.~\ref{sec:detectors}, before we give the details of our numerical simulations in Sec.~\ref{sec:numerics}. Section~\ref{sec:results} is devoted to the presentation of the outcome of our simulations and the corresponding discussion. Finally, in Sec.~\ref{sec:conclusions}, we briefly summarize the results and present our conclusions.

\section{Background physics}
\label{sec:background}

Indirect neutrino signals from DM annihilations in the Sun and the Earth have been studied extensively in the literature~\cite{Silk:1985ax,Krauss:1985ks,Gaisser:1986ha,Blennow:2007tw,Bueno:2004dv,Mena:2007ty,Kumar:2011hi,Agarwalla:2011yy,Desai:2004pq,Niro:2009mw,Kappl:2011kz,Boliev:1995xz,Barger:2001ur,Ackermann:2005fr,Halzen:2005ar,Ritz:1987mh,Gould92,Jungman:1995df,Bergstrom:1996kp,Bergstrom:1998xh,Abbasi:2009uz,Crotty:2002mv,Hooper:2002gs,Cirelli:2005gh,Lehnert:2007fv,Barger:2007xf,Blennow:2009ag,Esmaili:2009ks,Esmaili:2010wa,Das:2011yr,Farzan:2011ck,Silverwood:2012tp,Kumar:2012uh,Boliev:2013ai}, in particular in connection to neutrino telescopes. The general idea is based on DM particles from the galactic DM halo scattering on the solar matter and losing enough energy to become gravitationally bound to the Sun. The DM then loses more and more energy in subsequent interactions before finally thermalizing in the center of the Sun, where they can annihilate with other DM particles into SM ones. Since neutrinos are the only SM particles that can escape the solar interior, these annihilations can be probed by searching for the resulting neutrinos at the end of the decay chain.

The DM cross-section on SM matter can be effectively divided into a spin dependent~(SD) and a spin independent~(SI) cross-section. The latter of the two scales as $A^2$, where $A$ is the number of nucleons within the nucleus on which DM scatters, due to constructive interference among the scatters on individual nucleons. The SD cross-section, on the other hand, depends on the overall spin of the nucleus. Therefore, the bounds on the SI cross-sections from direct detection experiments tend to be much stronger than the corresponding bounds on SD cross-sections, since typically nuclei with a large number of nucleons are exploited. As the Sun consists mostly of elements with a small number of nucleons, the SD cross-section instead tends to dominate and it is mainly this type of cross-section that neutrino experiments will be able to probe. We therefore focus on the SD cross-section throughout this study.

The typical capture rate of DM particles in the Sun by a species $i$ can be described by the relation~\cite{Gould92,Wikstrom:2009kw}
\begin{equation}
C_i = \int dV\ \frac{\rho_\DM \rho_{i}(r)}{2m_\DM \mu_i^2} \sigma_i \int du\ \frac{f(u)}u \int dE |F(E)|^2,
\end{equation}
where $\rho_\DM$ is the local DM density, $\rho_i(r)$ is the density of the species $i$ at a distance $r$ from the Sun's center, $m_\DM$ is the dark matter mass, $\mu_i$ the reduced mass of the DM-$i$ system, $f(u)$ the velocity distribution function for DM, which we will assume Maxwellian, and $F(E)$ is a form factor for the scattering on $i$ as a function of the nuclear recoil $E$, which is essentially 1 for the dominant scatter in hydrogen at all relevant energies. For details on this procedure, see~\cite{Wikstrom:2009kw,Kappl:2011kz}. For the purpose of our study, the main thing to notice is that the capture rate is directly proportional to the capture cross-section $\sigma_i$. Taking only capture and annihilation into account, the number of DM particles in the Sun follows the differential equation
\begin{equation}
 \frac{dN}{dt} = \sum_i C_i - C_A N^2 \equiv C_C- C_A N^2,
\end{equation}
where $C_A$ is a constant describing the rate of annihilations between two DM particles. Since every annihilation involves two DM particles, the annihilation rate $\Gamma_A = C_A N^2/2$. In the steady state limit, where $dN/dt = 0$, we have detailed balance between capture and annihilation
\begin{equation}
 \Gamma_A = \frac 12 C_C.
 \label{eq:detailed}
\end{equation}
Thus, also the annihilation rate is directly proportional to the capture cross-section in this case. The the typical time-scale to reach the detailed balance is given by
\begin{equation}
 \tau_{\rm db} \sim \frac{1}{\sqrt{C_C C_A}}
\end{equation}
and the requirement that detailed balance is reached during the age of the Sun therefore translates to $C_C C_A \gtrsim (5\cdot 10^9\ {\rm years})^{-2}$. If $C_A$ is fixed such that the DM abundance is provided as a thermal relic, then this puts a lower limit on $C_C$ and thus also on $\sigma$, the DM-nucleon cross-section. For the masses treated in this study, the detailed balance starts being in question for cross-sections $\sigma \simeq 5 \cdot 10^{-3}$~fb. As we will see, such small cross-sections cannot be probed by the detectors in question, regardless of the annihilation channel, and we therefore assume that detailed balance holds.

Fixing $v_\odot = 220$~km/s, $\rho_\DM = 0.3$~g/cm$^3$, and using the BP2000 solar model~\cite{Bahcall:2000nu} for the composition of the Sun, the annihilation rate $\Gamma_{A,p}$ into $p\overline{p}$ is thus given by
\begin{equation}
 \Gamma_{A,p} = \BR(\DM\DMbar\to p\overline{p}) \Gamma_A = \BR(\DM\DMbar \to p \overline{p}) \sigma g(m_\DM) \equiv \br{p} g(m_\DM)
\end{equation}
where we have introduced the parameters $\br{p} = \sigma \BR(\DM\DMbar \to p \overline{p})$ and $g(m_\DM)$ is a function of the DM mass which can be computed from the expression for the capture rate.

In this study, we will be concerned with the annihilation rates into three different annihilation channels:
\begin{equation}
  \DM\DMbar \longrightarrow q\overline{q}, \quad \DM\DMbar \longrightarrow \tau^-\tau^+, \quad {\rm and} \quad \DM\DMbar \longrightarrow \nu\overline{\nu}.
\end{equation}
Here, $q$ denotes a generic charm or bottom quark channel, normalized to the annihilation into bottom quarks, while $\nu$ denotes a generic annihilation directly into any flavor neutrino. As demonstrated in \cite{Agarwalla:2011yy}, the neutrino spectra of the individual channels within these generic categories have negligible differences and the inclusion of separate degrees of freedom for each would only lead to almost perfect degeneracies among the components where only the following combinations can be independently constraint:
\begin{equation}
\br{q} = \br{b} + 0.4\ \br{c} \quad {\rm and} \quad \br{\nu} = \br{\nu_e} + \br{\nu_\mu} + \br{\nu_\tau}.
\end{equation}

We now have a four-dimensional parameter space containing $m_\DM$, $\br{q}$, $\br{\tau}$, and $\br{\nu}$, all of which are related to the properties of DM. Depending on the parameters, the resulting neutrino spectrum may have very different characteristics, from the highly peaked mono-chromatic neutrino channel to the very soft quark channel, and for a varying range of different DM masses. We therefore pose the question of how sensitive future neutrino detectors will be to these parameters, since determining them would give us important insight to the properties and nature of DM. This is a subject that has already been partially studied in several contributions to the literature~\cite{Mena:2007ty,Agarwalla:2011yy,Das:2011yr}, ranging from huge neutrino telescopes to the smaller scale neutrino detectors used in terrestrial experiments. However, a common approach within the literature has been to fix some of these parameters, usually the DM mass, while exploring the effect of the others. In this study, we extend these previous analyses to the whole parameter space to explore possible parameter degeneracies as well as the capability of these detectors to actually measure the DM mass on their own (without fixing it).

The most stringent upper bounds on the spin dependent cross-section of DM on protons in the mass region 10--25~GeV are given by the COUPP~\cite{Behnke:2012ys} and SIMPLE~\cite{Felizardo:2011uw} collaborations and are of $\mathcal O(10)$~fb. The limits from an analysis of the Super-K bounds~\cite{Kappl:2011kz} are around 0.2~fb for annihilations directly into neutrinos, 1~fb for annihilations into $\tau^+\tau^-$, and 10~fb for annihilations into $b\bar{b}$ (again fixing all other parameters). The values of the cross-sections that we will use as examples in our results section are all compatible with these limits. Note that, since a branching ratio can be at most one, if a signal is seen in the type of search discussed in this study, then the real capture cross-section must be as large or larger. At the same time, the cross-sections studied in \cite{Kappl:2011kz} assume a 100~\% branching into the respective channels and should therefore be treated as bounds on the very parameters that we are studying. Thus, $\br{\nu} = 0.1$~fb would not necessarily imply a large branching ratio into neutrinos, even though the quoted Super-K bound is 0.2~fb, since the actual capture rate may be as large as 10~fb.

\section{Detector technologies}
\label{sec:detectors}

Large underground laboratories have been essential for the detection and study of very low luminosity processes such as neutrino physics or proton decay due to the necessary overburden to avoid additional backgrounds induced by atmospheric muons. The proposed next generation of big underground neutrino detectors aims to continue and build upon the success of present experiments entering the 100--1000~\kton{} mass range. These detectors could play a key role in the future beam based oscillation experiments as well as serving as neutrino telescopes for astrophysics studies and increasing the sensitivity to proton decay. With the recent discovery of a large leptonic mixing angle $\theta_{13}$ by reactor neutrino experiments, this new generation of neutrino experiments could provide answers to fundamental questions about the neutrino mass hierarchy and leptonic CP-violation. With the increased angular and energy resolutions of these experiments, they will also provide a good opportunity for probing eventual neutrino fluxes from DM annihilations inside the Sun for DM masses in the $\mathcal O(10)$~GeV range, which cannot be reached by neutrino telescopes due to a too large spacing between the optical modules.

In this study, we focus on the liquid argon (LAr) time projection chamber and magnetized iron calorimiter (MIND) detector technologies. While water \v{C}erenkov and liquid scintillator technologies could also be of interest, their capabilities for angular and energy resolution drop for energies above the quasi-elastic regime.
For the detector technologies included in this study, we follow closely Ref.~\cite{Agarwalla:2011yy}, where the very same detector technologies were investigated.

\paragraph{Liquid Argon Time Projection Chamber:} The LAr TPC is a modern version of the bubble chamber. By applying an electric field to very pure argon, ionizing particles passing through the detector can free electrons which subsequently drift to the instrumented surface of the detector where a three-dimensional read-out can be performed. This detector technology has been developed largely by R\&D performed by the ICARUS collaboration~\cite{Amerio:2004ze}. It is under consideration as a detector for the future LAGUNA~\cite{laguna} and LBNE~\cite{lbne,Barger:2007yw} projects. LBNE is currently planned as a 10~\kton{} detector on the surface in the initial stages of the project. However, the planed next stage of the project consists of an upgrade to 34~\kton{} underground detector at a later time. Similarly, the first stage of the LAGUNA LAr detector would correspond to 20~\kton{} with the aim of a gradual upgrade up to 100~\kton~\cite{Stahl:2012exa}. Here we will consider the physics potential of these upgrades assuming a 34/100~\kton{} setup, keeping in mind that our results then scale in time, \emph{i.e.}, our assumed 10~years of running for the 34~\kton{} detector would correspond to 34~years of running for the 10~\kton{} detector (although this assumes that the detector is moved underground). Due to the difficulty in applying a magnetic field to the volume of liquid argon, we consider this detector type as being able to distinguish only the flavors of neutrinos while having no sensitivity to whether or not an interaction was caused by a neutrino or anti-neutrino.

\paragraph{Magnetized Iron Calorimeter:} The MIND~\cite{Bayes:2012ex} technology consists of iron modules with instrumented layers in between, where the energy deposit of charged particles can be measured. Since electrons and positrons will not pass through the iron, this technology is only sensitive to muon-type neutrinos. On the other hand, a volume of iron can be easily magnetized, providing sensitivity to distinguish neutrino from anti-neutrino events. A 100~\kton{} MIND detector is considered the baseline for implementation in a neutrino factory experiment. Currently, the India-based Neutrino Observatory (INO)~\cite{ino} detector is aiming for a MIND-type detector of 50~\kton{} (alternatively 100~\kton{} after upgrade) primarily in order to study atmospheric neutrinos but also as a possible far detector of a neutrino beam experiment. As for the LAr detectors, we will here study the capabilities of the fully upgraded 100~\kton{} version. The results would be equivalent to those for a 50~\kton{} detector with twice the running time.

For further details of our experimental setup, we refer the reader to \cite{Agarwalla:2011yy}, in particular \tab~I and \fig~I, where the characteristics of the detectors are described in detail.

\section{Numerical implementation}
\label{sec:numerics}

Our numerical implementation is an extension of that of \cite{Agarwalla:2011yy} in order to include the possibility of varying the DM mass within the simulations. We perform the following steps in order to accomplish this:
\begin{enumerate}
 \item {\bf Simulate the neutrino spectra.} We start by simulating neutrino spectra for all three annihilation channels using the WimpSim software~\cite{Blennow:2007tw,wimpsim}, which in turn is using the nusigma~\cite{nusigma}, DarkSUSY~\cite{Gondolo:2004sc}, and Pythia~\cite{Sjostrand:2006za} codes. While computing the spectra during the simulations for the specific masses involved would be preferable, this is too computationally expensive to be feasible and instead we simulate the spectra for masses between 5 and 30~GeV with a 0.5~GeV resolution and later interpolate between these once the results have been binned into 1~GeV bins. We chose 0.5~GeV in order to have a manageable amount of data while still having a resolution smaller than the bin size for better interpolation. 
For the direct annihilation into neutrinos, we assume an equal branching into the different neutrino channels in order to suppress oscillation effects on the monochromatic peak in this flux~\cite{Blennow:2007tw,Esmaili:2009ks,Esmaili:2010wa,Farzan:2011ck}, which could otherwise interfere with the results due to the solar oscillation length being of the same order of magnitude as the difference in the Sun-Earth distance between aphelion and perihelion for neutrinos with energies $\mathcal O(20)$~GeV. The resulting fluxes would thus be strongly dependent to a change in the neutrino mass squared difference $\Delta m_{21}^2$ within its current errors.
For the other simulations, we used the neutrino oscillation parameters summarized in \tab~\ref{tab:oscparams} in order to compute the fluxes. Since the quark and tau channels give rise to continuous fluxes rather than monochromatic ones, the energy resolutions of the detectors will give an averaged effect. It should be noted that the exact values of the neutrino oscillation parameters are not of crucial importance here, as oscillations mainly act to equilibrate the fluxes among the neutrino flavors for continuous spectra~\cite{Blennow:2007tw}. We therefore limit this study to a fixed set of oscillation parameters in the normal neutrino mass hierarchy.
 \begin{table}
 \begin{center}
  \begin{tabular}{|c|c||c|c||c|c|}
  \hline
   \multicolumn{2}{|c||}{\bf Angles [$^\circ$]} & \multicolumn{2}{|c||}{\bf Mass splittings [eV$^2$]} & \multicolumn{2}{|c|}{\bf Phase} \\
   \hline
   \hline
   $\theta_{12}$ & 33.65 & $\Delta m_{21}^2$ & $7.54\cdot 10^{-5}$ & $\delta$ & 3.39 \\
   \hline
   $\theta_{23}$ & 38.41 & $\Delta m_{31}^2$ & $2.39\cdot 10^{-3}$ \\
   \cline{1-4}
   $\theta_{13}$ & 8.93  \\
   \cline{1-2}
  \end{tabular}
 \end{center}
 \caption{The oscillation parameters used in our simulations. The mixing angles and mass square differences have been chosen to correspond to the best-fit values of \cite{Fogli:2012ua}. Since the global errors are small for our purposes and the result is insensitive to the precise values of the parameters, we do not study variations of the parameters in this study. The same is true for changing from inverted to normal mass hierarchy. \label{tab:oscparams}}
 \end{table}
 \item {\bf Simulate the detector response.} The fluxes from WimpSim were then used as input to the GLoBES software~\cite{Huber:2004ka,Huber:2007ji} in order to simulate the detector response. We bin the events in energy into 28 bins of size 1~GeV from 2~GeV to 30~GeV. The GLoBES flux normalization was cross-checked against a simple integration for exact energy resolution and 100~\% efficiency in order to obtain the correct event rates. Finally, the detector resolution and efficiencies were switched on and the event rates were extracted for a capture cross-section of 1~fb and an exposure of 1~year. The detector response for different cross-sections and exposures were then obtained by simply scaling the response.
 \item {\bf Simulate the background.} The main background to our signal processes is that of atmospheric neutrinos interacting in the detector. As atmospheric neutrinos come from all directions, this background can be reduced by an angular cut. For this purpose, we use the atmospheric neutrino fluxes of \cite{Honda:2011nf} and oscillate them with the same oscillation parameters as those used for the signal simulation. For the same energy binning as in the previous step, we compute the atmospheric background in the detector. This is done by taking a sky average and then only including events within an angle from the Sun of the typical angular resolution of the detector. The assumptions for the angular resolution of each detector as a function of the energy are summarized in Fig.~1 of Ref.~\cite{Agarwalla:2011yy}. We use the fluxes at the Frejus site from~\cite{Honda:2011nf} as representative when simulating the atmospheric background. As for the signal, the atmospheric background rate is computed for one year of running and later rescaled to the appropriate running time.
 \item {\bf Implementation into a Markov Chain Monte Carlo.} By interpolating the binned spectra, we can now obtain an approximation for the expected spectrum at an arbitrary DM mass. We can thus perform a Markov Chain Monte Carlo~(MCMC) scan of the parameter space for any given parameter values within our restricted mass range. We perform simulations for the detectors outlined in the previous section and varying assumptions on the branching ratios and $m_\DM$. The MCMC scan is implemented through the MonteCUBES software~\cite{Blennow:2009pk}, which employs the Metropolis-Hastings algorithm in order to sample the posterior distribution. This sampling of the posterior distribution is then used to construct credible regions through the use of the MonteCUBES Matlab GUI. For each MCMC scan, we have generated a total of four chains with $2\cdot 10^5$ samples each in order to have a reasonably well sampled posterior distribution. The convergence was good in all of the simulations presented in this study ($R < 1.001$~\cite{Gelman:1992zz}). Our MCMC simulations always assume a running time of 10 full years.
\end{enumerate}

\section{Results}
\label{sec:results}

\subsection{Detector rates}
\label{sec:detectorrates}

\begin{figure}
\begin{center}
\includegraphics[width=0.9\textwidth]{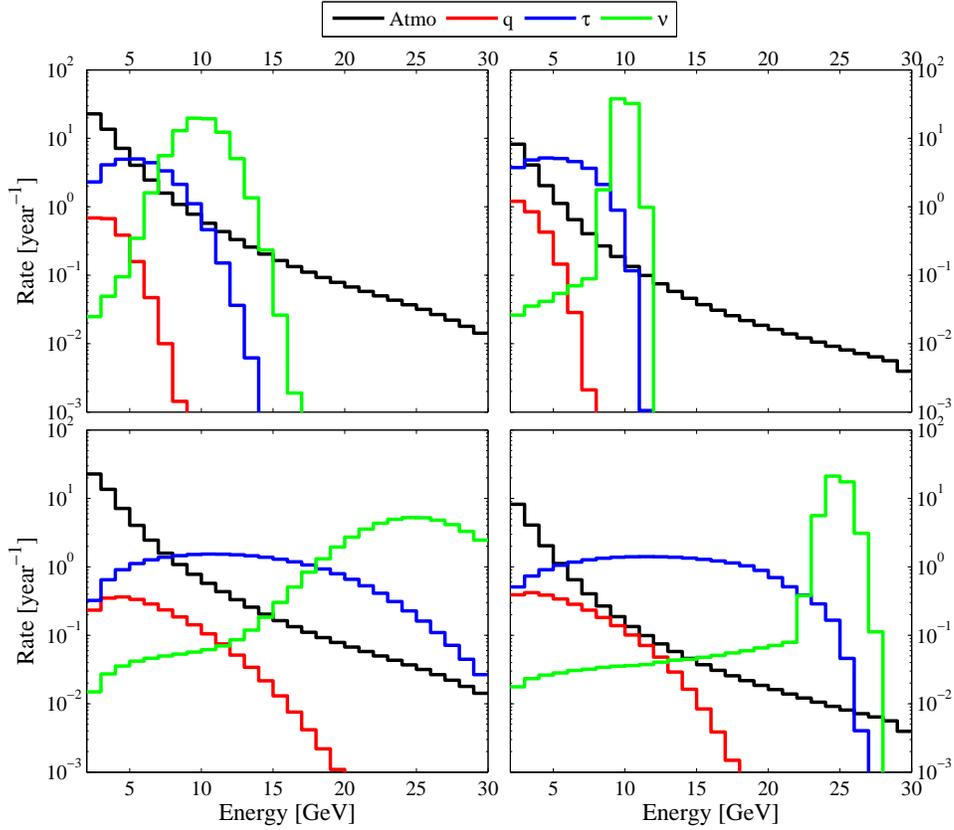}
\end{center}
\caption{The expected muon event rates in the different detector types for the different annihilation channels under study. The left (right) panels correspond to a MIND (LAr) detector of mass 100~\kton{}. The assumed mass of the DM particles is 10~GeV (25~GeV) in the upper (lower) panels. The fluxes are produced assuming $\br{x} = 1$~fb and full branching into the channel. Since LAr detectors do not have charge identification, we show the sum of $\mu^-$ and $\mu^+$ events for the plots using LAr technology, while the plots for the MIND detectors are for $\mu^-$ events only. \label{fig:rates}}
\end{figure}

In \fig~\ref{fig:rates}, we present the full binned spectrum in the energy range 2--30~GeV for the different annihilation channels assuming a 100~\kton{} detector and a 1~fb capture cross-section. As can be seen from this figure, the neutrino annihilation channel offers the sharpest peak in the spectrum and is therefore naturally expected to give the best sensitivity to the DM mass. Furthermore, the peak of the neutrino channel coincides with the DM mass. The $\tau$ channel spectrum is broad and hard enough to be visible above the atmospheric background rate and can also be exploited to provide some sensitivity to the DM mass. Finally, the softer and weaker quark channel spectrum will have a harder time providing a good enough signal-to-background ratio anywhere in the spectrum and the DM mass determination using the spectral cut-off will be very difficult unless capture cross-sections larger than 1~fb are invoked.
Furthermore, in all cases, the largest expected event rates reach a few per year, indicating that a full run of $\mathcal O(10)$~years will be necessary to obtain reasonable statistics.

\subsection{Mass determination}

\begin{figure}
\begin{center}
\includegraphics[width=0.48\textwidth]{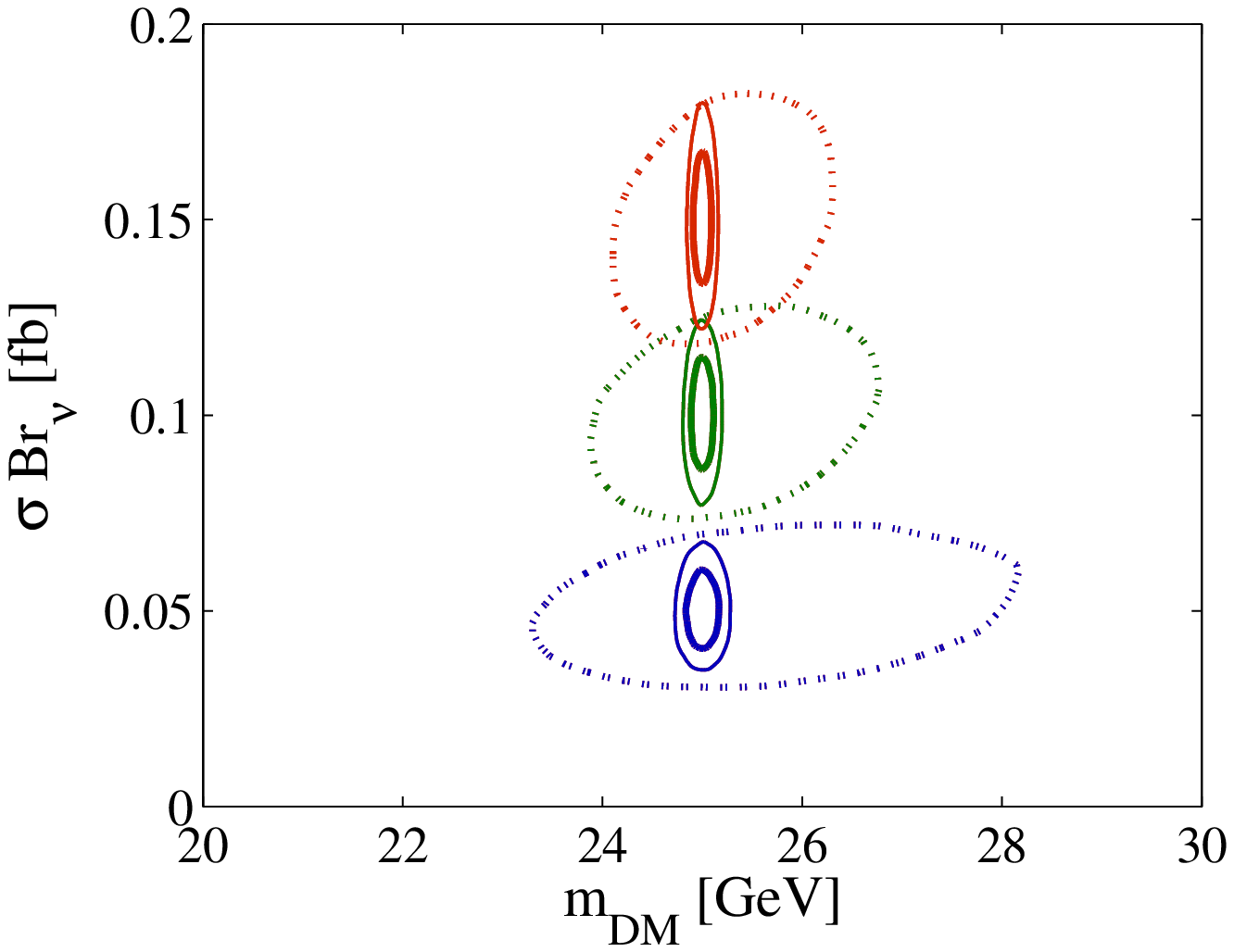}
\includegraphics[width=0.48\textwidth]{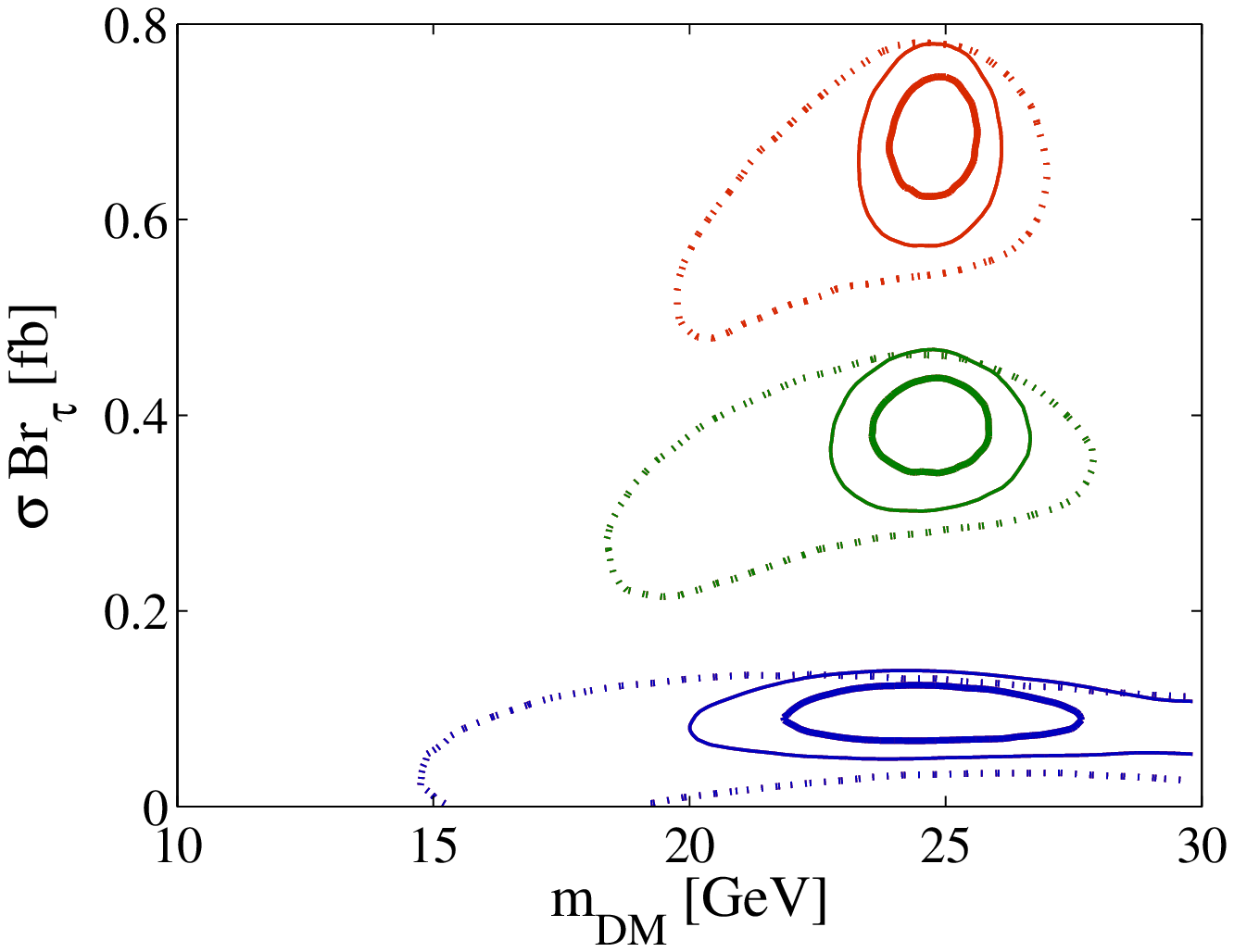} \\
\includegraphics[width=0.48\textwidth]{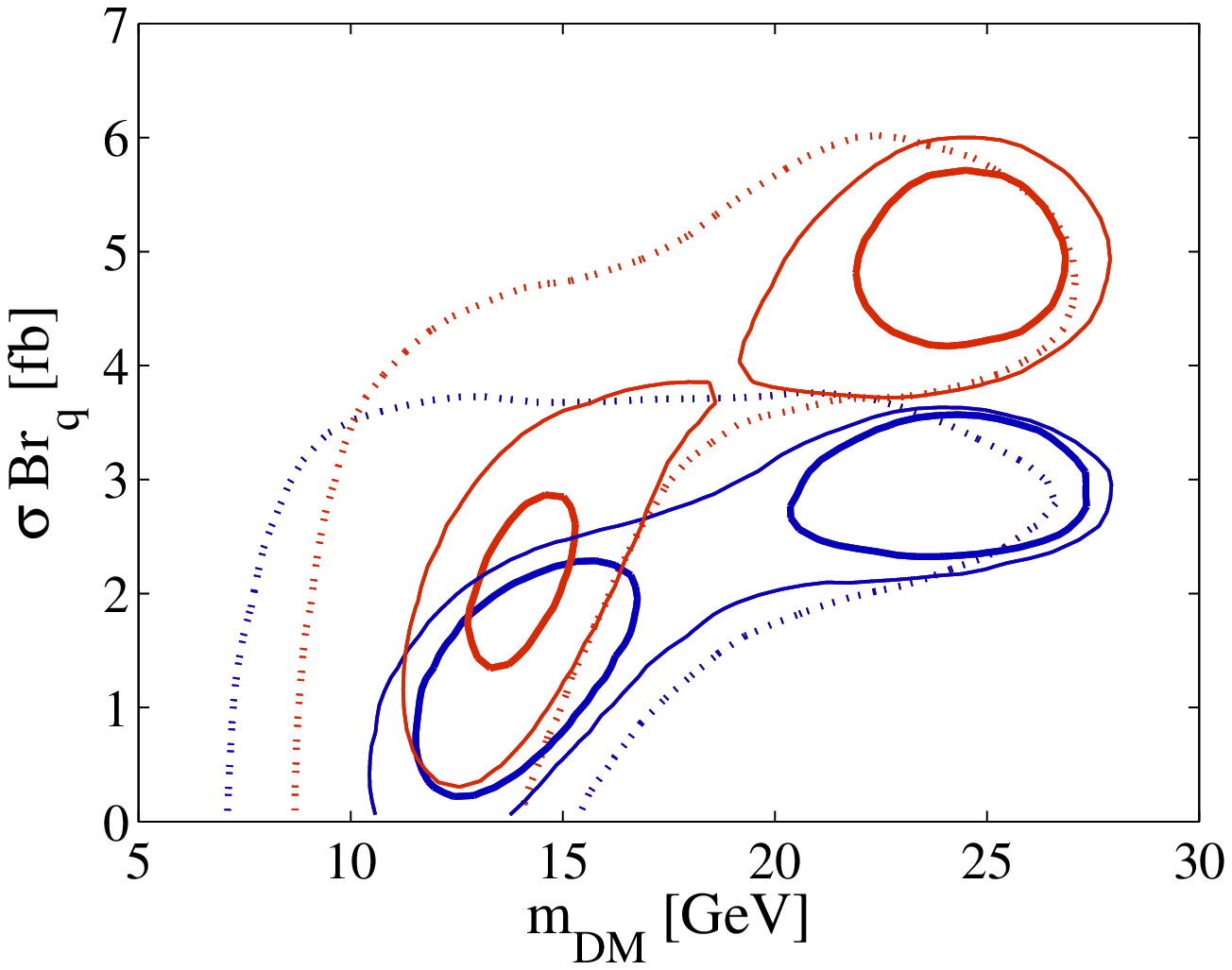}
\caption{\label{fig:mBr} The 90~\% credible regions for the mean experimental outcomes for annihilations of $m_\DM = 25$~GeV DM into neutrinos (upper left panel), taus (upper right panel), and quarks (bottom panel). In all panels, the thick (thin) solid lines correspond to the 100~kt (34~kt) LAr detector, while the dotted lines correspond to the 100~kt MIND detector. For the upper two panels, different lines for the same experiment correspond to different values for $\br{\nu/\tau}$: $\br{\nu} = 0.05$~(blue), 0.1~(green), and 0.15~fb~(red) for the left panel and $\br{\tau} = 0.1$~(blue), 0.4~(green), and 0.7~fb~(red) for the right. Due to the significant overlap in the lower panel, we show the results only for $\br{q} = 3$~(blue) and 5~fb~(red).}
\end{center}
\end{figure}

In \fig~\ref{fig:mBr}, we show the reconstructed 90~\% credible regions for the mean outcomes of the experiments for a DM mass of 25~GeV and different annihilation channels. For each channel, we show examples for three different values of $\br{x}$, except for the case of annihilation into quarks, where we show only two values in order to keep the figure reasonably clean. As already anticipated, we find that the hardest channels provide the best DM mass determination.
As can be seen from the two upper panels of the figure, the high energy resolution for the LAr detectors typically give good results for determining the DM mass for the neutrino and tau channels, providing error bars of around 1~GeV for the tau channel which improve by almost an order of magnitude for the neutrino channel.
For the worse energy resolution assumed for the MIND detector the reconstructed error bars increase, with 1~GeV for the neutrino channel with the largest cross section and presenting significantly worse mass determination compared to the LAr detectors in all cases. The increased mass sensitivity for increased $\br{x}$ is also naturally expected, as a higher number of events can be accumulated. In the opposite end, when $\br{x}$ is close to the sensitivity limit of the experiment, very little sensitivity to the mass is achieved, given the low statistics and poor signal-to-background ratio. 

\begin{figure}
\begin{center}
\includegraphics[width=0.48\textwidth]{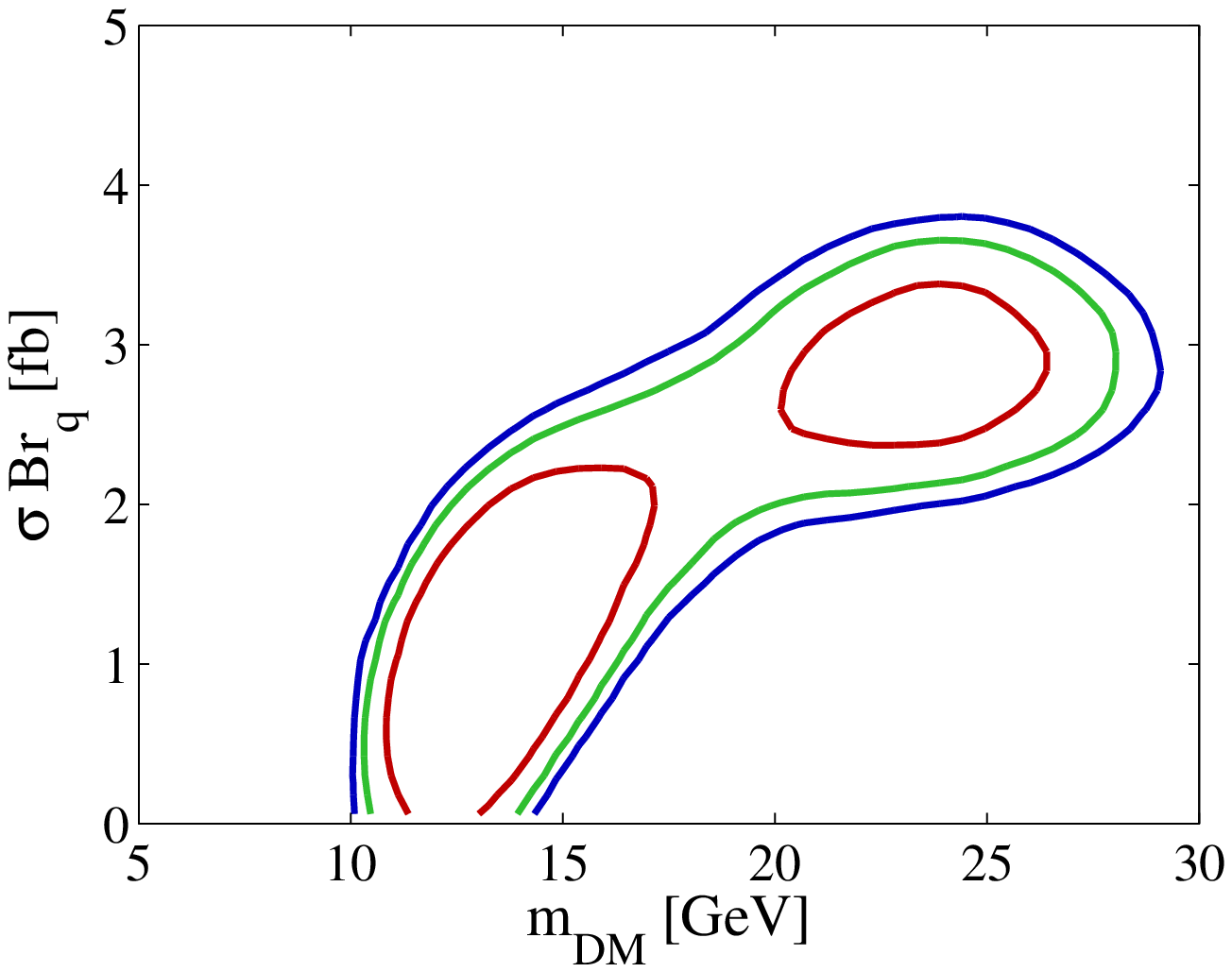}
\includegraphics[width=0.48\textwidth]{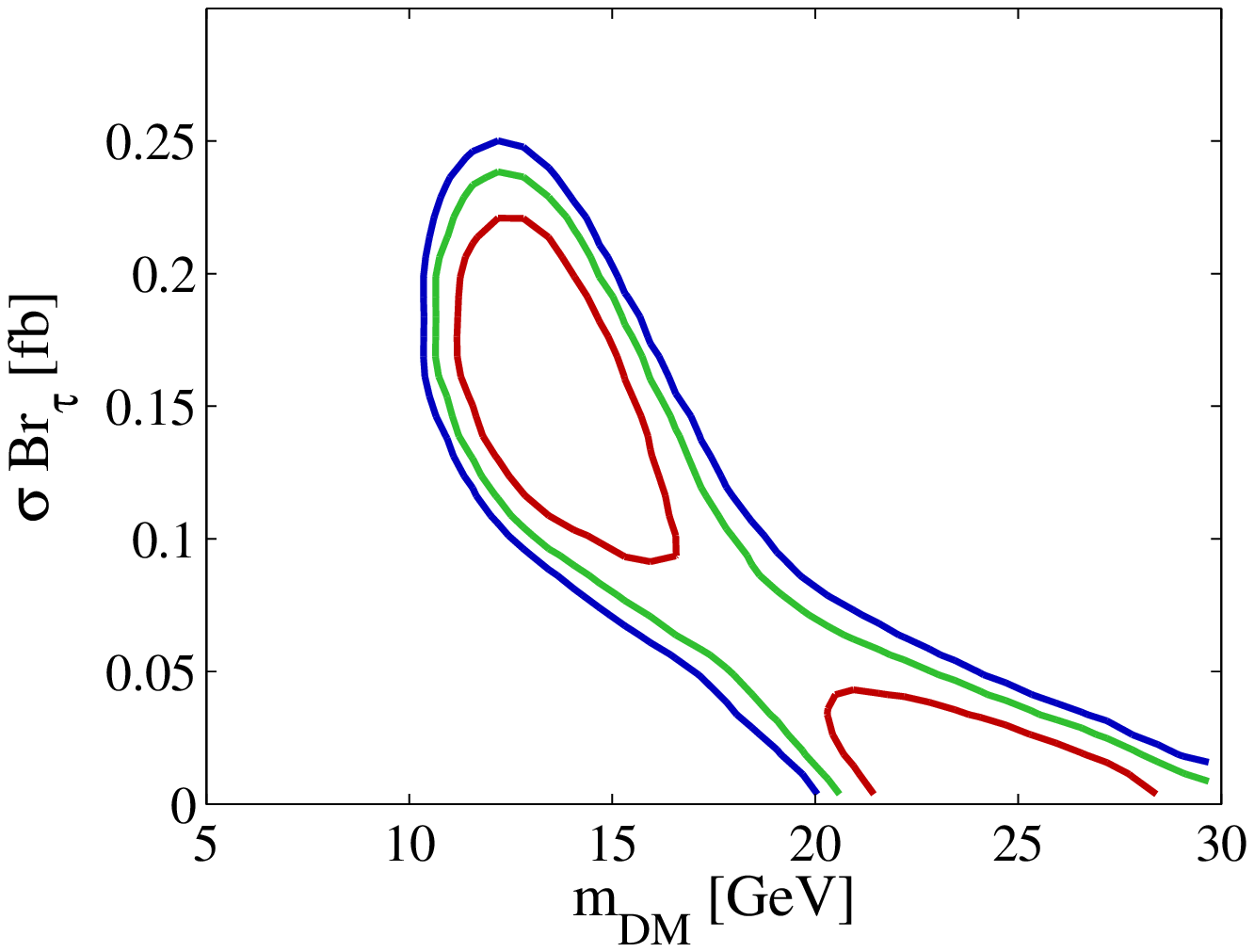} \\
\includegraphics[width=0.48\textwidth]{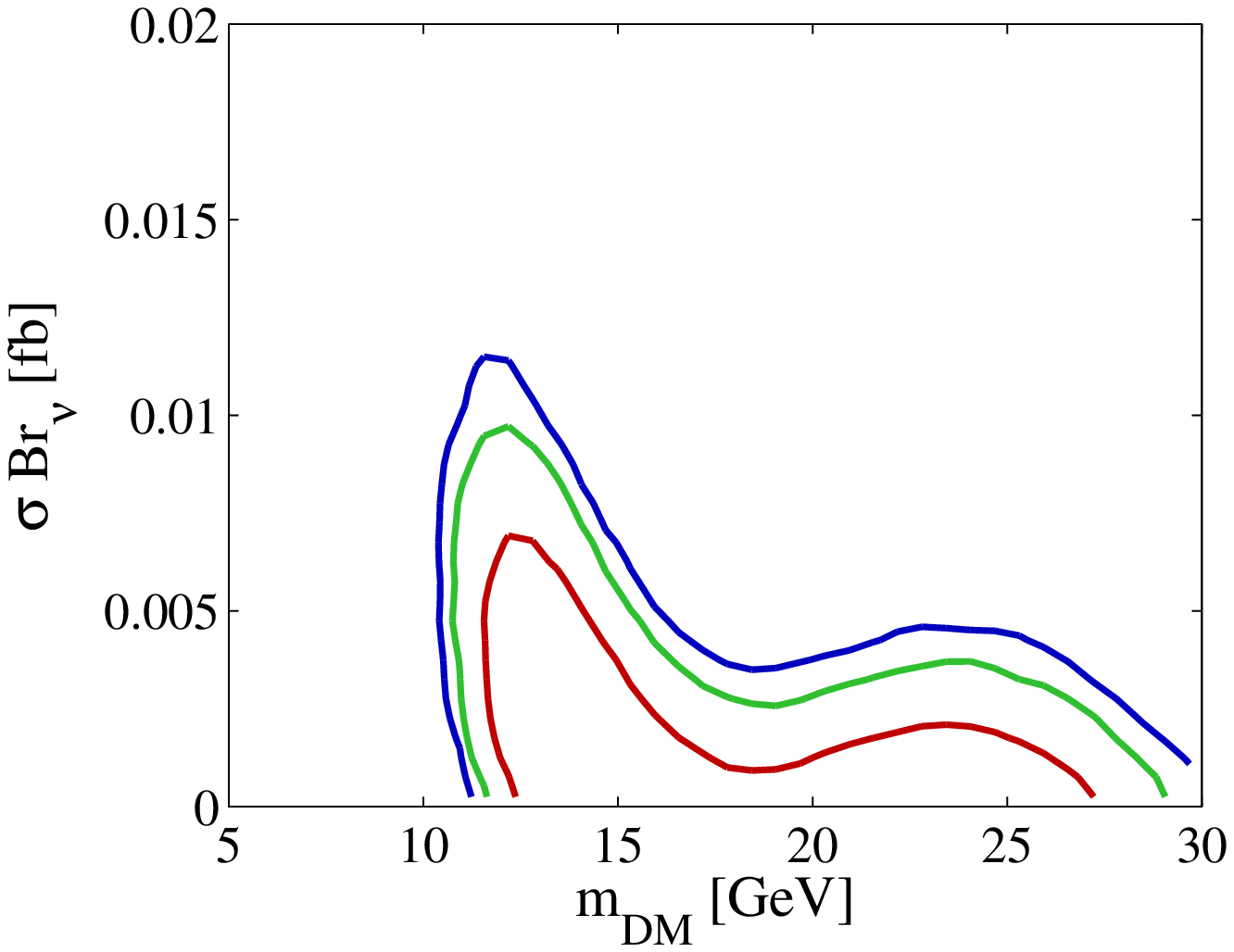}
\includegraphics[width=0.48\textwidth]{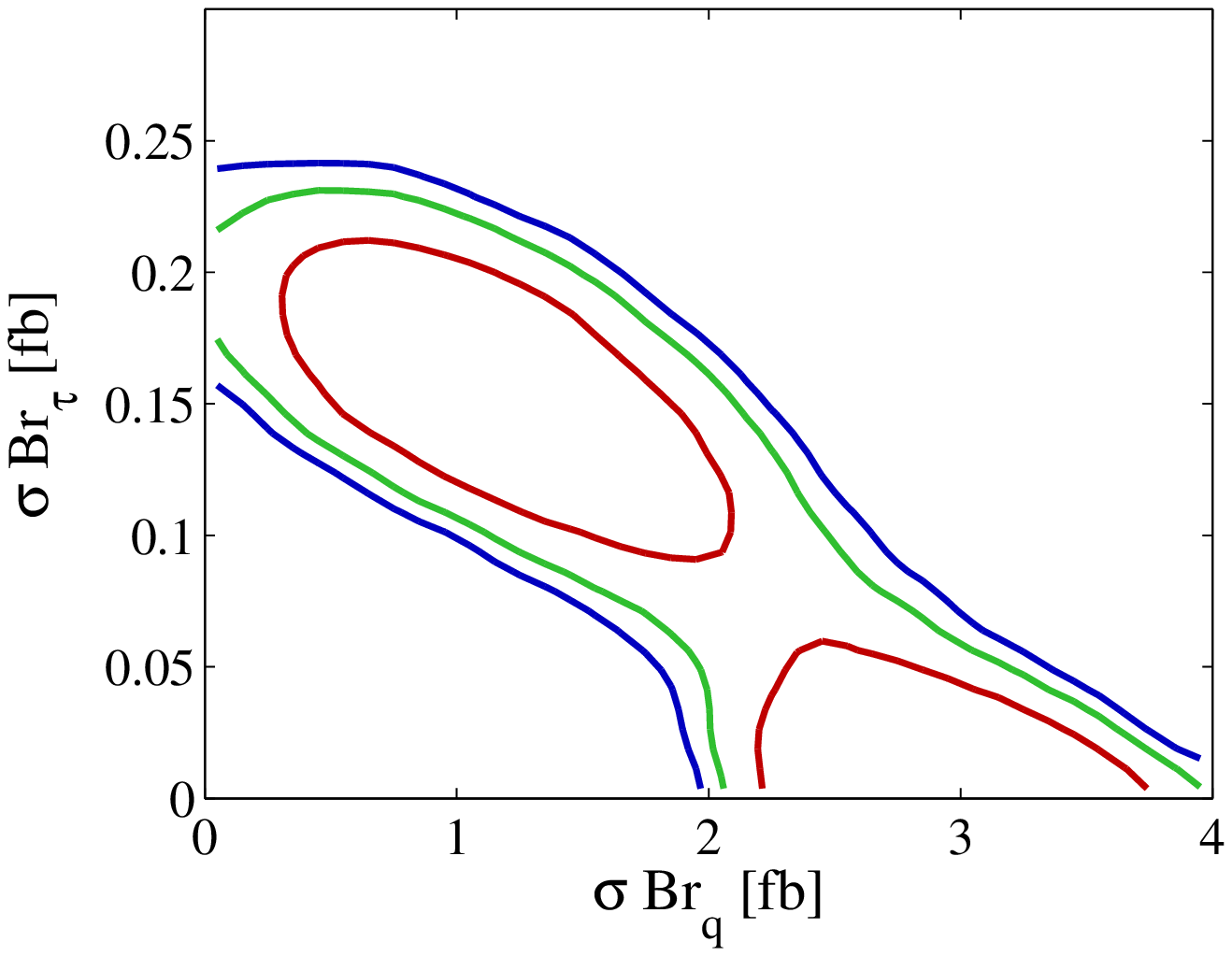}
\caption{\label{fig:deg} Two-dimensional credible regions at 68~\% (red), 90~\% (green), and 95~\% (blue) for the 34~kton LAr detector, $m_\DM = 25$~GeV and $\br{q} = 3$~fb.}
\end{center}
\end{figure}

\begin{figure}
\begin{center}
\includegraphics[width=0.6\textwidth]{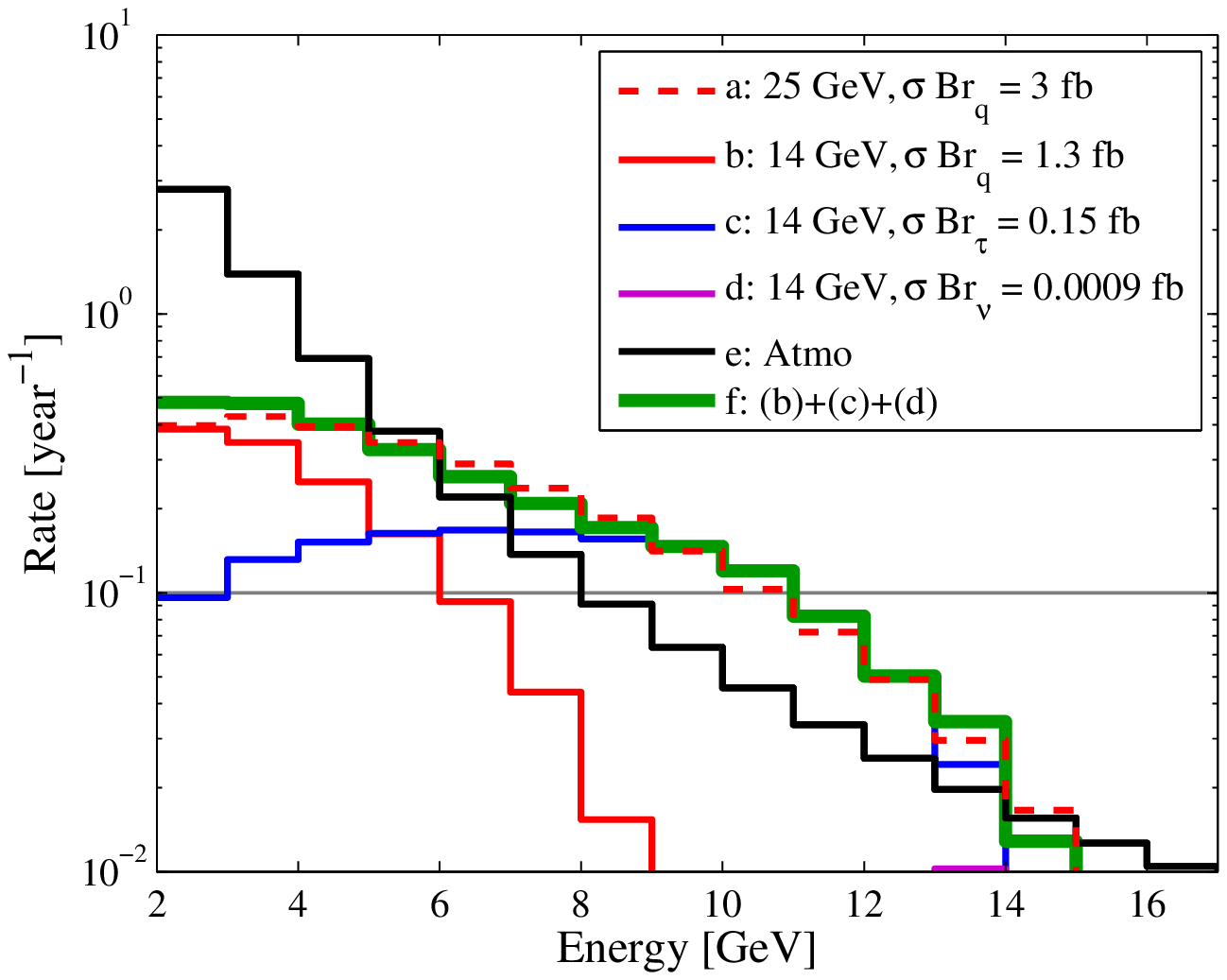}
\caption{\label{fig:degflux} The event rate from annihilations of DM with $m_\DM = 25$~GeV and $\br{q} = 3$~fb in a 34~\kton{} LAr detector is shown (red dashed line) together with the event rates from the degenerate solution $m = 14$~GeV, $\br{q} = 1.3$~fb~(red solid line), $\br{\tau} = 0.15$~fb~(blue solid line), and $\br{\nu} = 9.0\cdot 10^{-4}$~fb (purple solid line), and their sum (green thick line). Also shown is the atmospheric background rate (black solid line), while the horizontal gray line corresponds to an average of one event per bin over a 10~year period.}
\end{center}
\end{figure}

The lower panel of \fig~\ref{fig:mBr}, corresponds to the case in which DM is annihilating purely into quarks. As expected, the softer spectrum significantly hinders the reconstruction of the DM mass. Moreover, the plots also show clear signs of degeneracy for the $\br{q}$ channel at lower DM masses than the 25~GeV assumed, which further decreases the ability to accurately determine the mass.
In order to understand the origin of this degeneracy, in \fig~\ref{fig:deg} we show the credible regions for several of the projections for the case of $\br{q} = 3$~fb at the 34~kton LAr detector.
While this is enough in order to detect that there is a signal, it is not clear what the signal consists of. In particular, a strong degeneracy appears around $m_\DM \simeq 14$~GeV. The existence of this degeneracy was confirmed by using the built-in degeneracy finder (see \cite{mcbManual}) of MonteCUBES for all our experimental setups and for those masses where the degenerate solution could also be found within the region of study (10--25~GeV). For the case presented in \fig~\ref{fig:deg}, the best-fit point in the degenerate likelihood maximum is given by: $m_\DM = 14.2$~GeV, $\br{q} = 1.3$~fb, $\br{\tau} = 0.15$~fb, and $\br{\nu} = 9.0\cdot 10^{-4}$~fb. In order to visualize the origin of the degeneracy, we show in \fig~\ref{fig:degflux} the flux due to $\br{q} = 3$~fb for $m_\DM = 25$~GeV along with the different contributions to the flux from the degenerate solution and their sum.
From this figure it is clear that, in the energy range where the signals are well above the background level and at the same time have an event rate which could be accessible within a 10~year time frame, the $\br{q}$ and $\br{\tau}$ components of the degenerate solution add up to form a spectrum which is indeed very similar to that of the original input. For the degenerate solution, the part of the spectrum in the $\br{q}$ channel that is lost by going to a lower DM mass is compensated by the harder spectrum of the $\br{\tau}$ channel. This corresponds very well to what we would expect from \fig~\ref{fig:deg}, where we can see that a significant $\br{\tau}$ is always needed for the degenerate solution. At the same time, we can see that, as we decrease the mass towards that of the degenerate solution, $\br{q}$ decreases while $\br{\tau}$ increases correspondingly. It should also be noted that for the lightest DM masses in the degenerate solution, $\br{\nu}$ increases significantly. This is due to the $\tau$ spectrum not being hard enough at the low DM mass to reproduce the original signal without involving also the $\nu$ channel in the observable bins with the highest energy.

A similar degeneracy can be observed between a pure $\br{\tau}$ signal at lower masses and a combined degenerate signal dominated by $\br{q}$ at higher masses. However, this degeneracy is not as strong as that for the pure $\br{q}$ signal, as it is slightly more difficult to combine the different signals to reproduce the harder $\br{\tau}$-like spectrum.

\subsection{Sensitivity to different branching ratios}

\begin{figure}
\begin{center}
\includegraphics[width=0.48\textwidth]{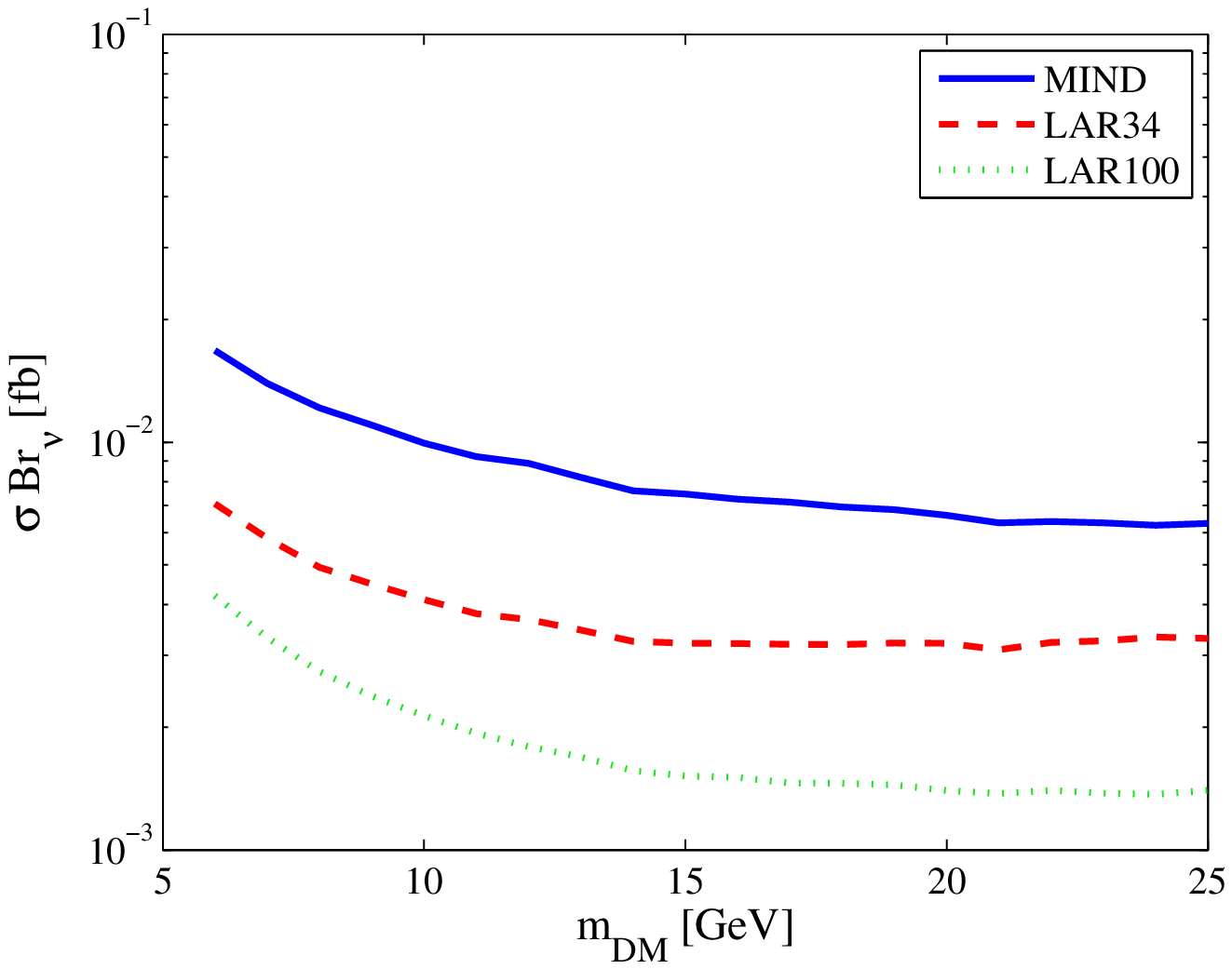}
\includegraphics[width=0.48\textwidth]{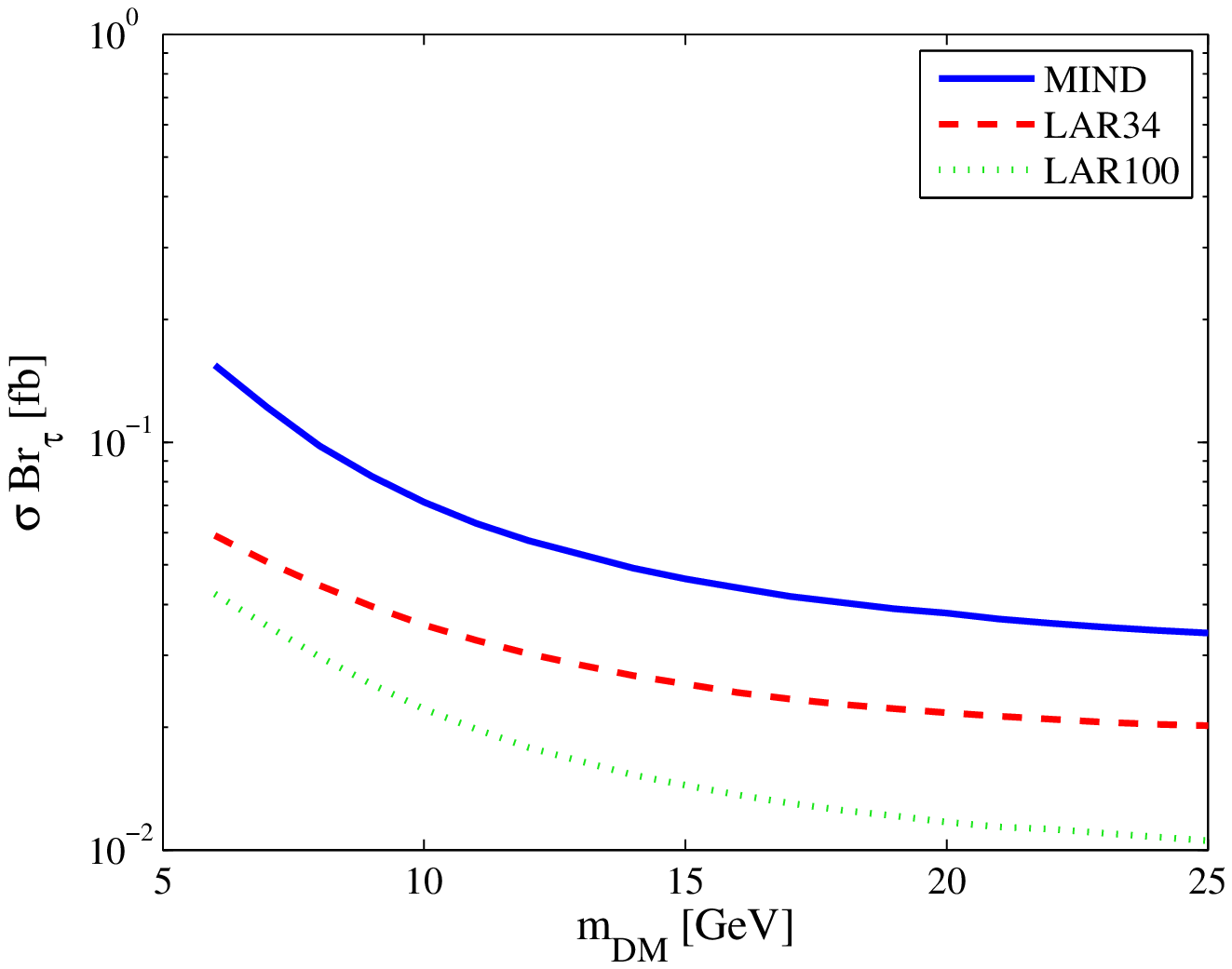} \\
\includegraphics[width=0.48\textwidth]{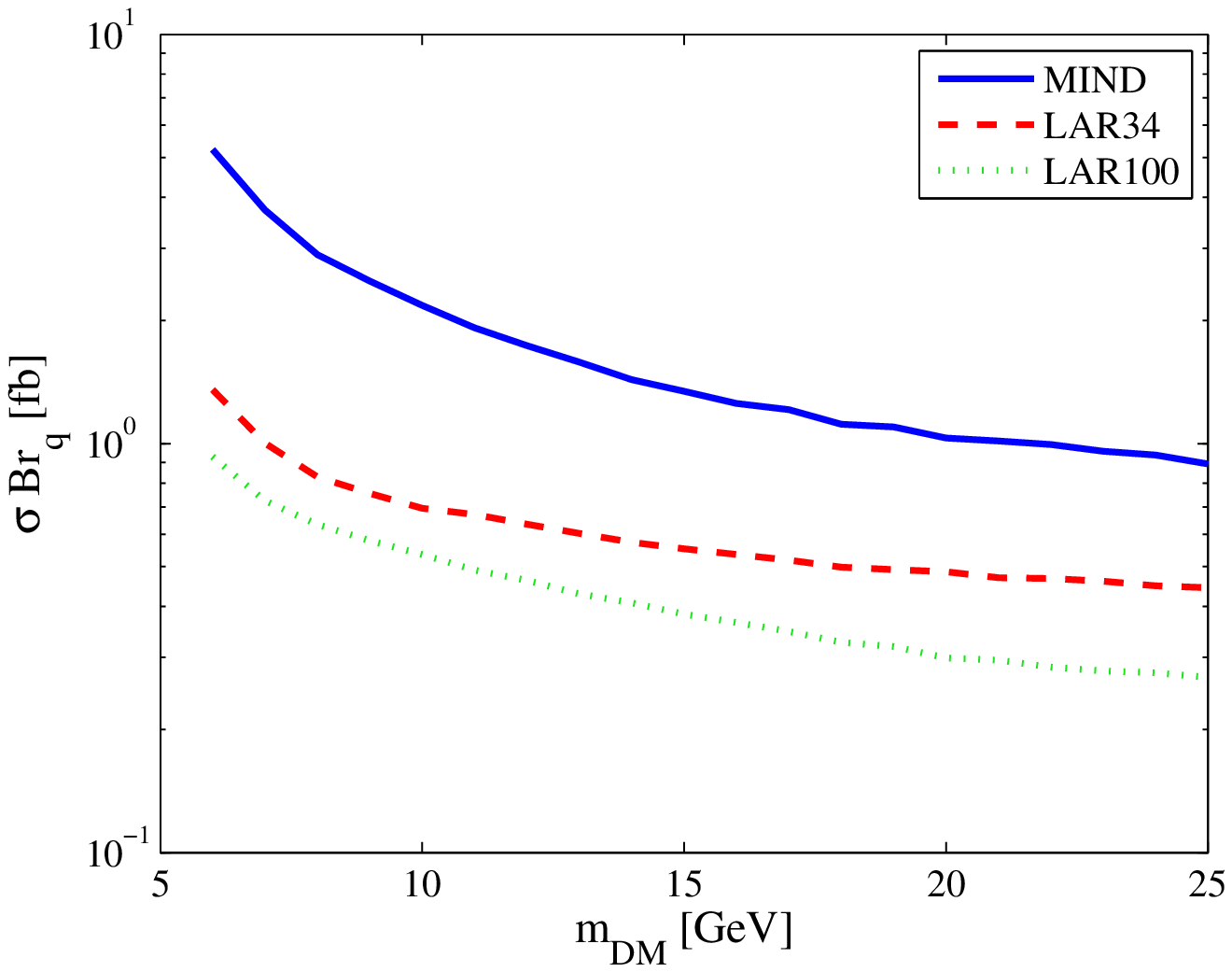}
\caption{\label{fig:massSens} The 90~\% credible upper bounds for the $\br{x}$ parameters for the $\nu$~(upper left), $\tau$~(upper right), and $q$~(lower) channels under the assumption of a mean experimental outcome in absence of signal as a function of the DM mass $m_\DM$, which has been assumed to be known.}
\end{center}
\end{figure}

In \fig~\ref{fig:massSens} we show the sensitivity of the different detectors to the various annihilation channels as a function of the DM mass. In order to derive these bounds, we compute the 90~\% credible upper bound on the different branching ratios in absence of signal. In these simulations the DM mass is kept fixed to the corresponding value in the horizontal axis
in order to obtain the sensitivity for that particular mass and to be comparable with bounds in other studies.
All other parameters have been marginalized over. While we have seen that the determination of the DM mass can be a challenging task in most of the cases presented above, it may still be of interest to pose the question of how well the different branching ratios can be constraint if the DM mass is known from a different source, such as direct or collider DM searches.
As expected, we see that for all cases the sensitivity becomes significantly worse at low DM masses, a result both of the signal appearing in a smaller energy range and the larger atmospheric background at low energies. For all channels, we can observe that the sensitivity of the 100~\kton{} LAr detector would be roughly half an order of magnitude (a factor of $\sim 3$) better than that of the 100~\kton{} MIND, with the largest difference in the neutrino channel, where the better energy resolution of the LAr detector at higher energies plays a decisive role. As expected, the sensitivity to the branching ratios becomes better the harder the spectrum for the channel is, with the sensitivity to $\br{\nu}$ ranging between $10^{-3}$ to $10^{-2}$~fb, the sensitivity to $\br{\tau}$ between $10^{-2}$ and $10^{-1}$~fb, and the sensitivity to $\br{q}$ between $0.3$ and 5~fb.

\section{Summary and Conclusions}
\label{sec:conclusions}

We have examined the neutrino signal from DM annihilations inside the interior of the Sun in different large projected neutrino detectors using the liquid argon (LAr) and magnetized iron (MIND) technologies. Such detectors are being discussed for the next generation of far detectors for neutrino beam oscillation experiments as well as for neutrino telescopes and, as we have shown, would also allow to probe the nature of DM. The energy reach of such detectors limits our study to relatively low DM masses $\mathcal O(10)$~GeV, which is quite complementary to similar studies in neutrino telescopes such as IceCube that operate at higher energies $\mathcal O(100-1000)$~GeV, but have limited sensitivity at lower energies due to a large distance between the optical modules. Our results are based on the detectors taking data for 10 full years.

In this study, we have extended the work in the existing literature by allowing the DM mass to take varying values when performing our fits. Studies similar to ours have been presented before in the literature, but typically fixing the DM mass at particular values. Thus, we have been able to study how well the DM mass, along with the branching ratios of the annihilation, can be determined from the shape of the neutrino spectrum.

We have shown that the best annihilation channel for a precise determination of the DM mass is the direct annihilation of DM into neutrinos, presenting us with an essentially monochromatic spectrum. The energy resolution of the detector is then of crucial importance for how well the DM mass can be determined, with error bars of $\mathcal O(0.1)$~GeV for the LAr type detectors and a few GeV for MIND (at a DM mass of 25~GeV). For the relatively hard spectrum of DM annihilations into $\tau^-\tau^+$ pairs, we still have a fair chance of constraining the DM mass, at least for the cases where the branching ratio is sufficiently large. The LAr detectors now show a mass resolution of a few GeV at best~(again for a DM mass of 25~GeV), while the MIND is again significantly worse. The worst sensitivity to the DM mass is found for the annihilation into quarks, which produces a very soft spectrum of neutrinos. Not only would this channel produce less neutrinos in total per annihilation, but it would also be challenged by a higher atmospheric background at low energies. In addition, the spectrum is so soft that the maximum energy bins for which events could be expected over a 10 year period (a rate larger than 0.1~events/year) do not correspond to the kinetic cutoff at the DM mass. The result is that the resolution for the mass suffers even for the LAr detectors.

In addition to these effects, we have also shown the existence of a degeneracy in the determination of the DM mass and the branching ratios. This degeneracy originates in the fact that the harder spectrum for annihilations into $\tau^-\tau^+$ pairs for a lower DM mass can be quite similar to that of annihilations into quarks for a higher DM mass. While this does not pose a real problem for the actual indirect detection of the neutrino signal from DM annihilations in the Sun (there is a clear signal over the atmospheric background), it would do so for the measurement of the DM mass and branching ratios and therefore for the determination of the nature of DM. It should be noted that if the DM mass (or branching ratios) would be determined from a different source, this degeneracy would be lifted.

In conclusion, the large future neutrino detectors intended as far detectors of neutrino beam experiments also present an exciting opportunity to learn more about the properties of DM. In particular, the DM mass can be probed at these detectors, but its accurate determination strongly depends on the actual branching ratios of DM annihilations and, when all parameters are simultaneously taken into account, strong degeneracies may appear.

\begin{acknowledgments}
We would like to thank Olga Mena for useful discussions. EFM acknowledges financial support by the European Union FP7 ITN INVISIBLES (Marie
Curie Actions, PITN-GA-2011-289442); the Spanish MINECO 
through the project FPA2009-09017 and the Ram\'on y Cajal programme (RYC-2011-07710)  
and the Comunidad Aut\'onoma de Madrid
through the project HEPHACOS P-ESP-00346.
\end{acknowledgments}

%\bibliographystyle{apsrev}
%\bibliography{refs}

%
%%%%%%%%%%%%%%%%%%%%%%%%%%%%%%%%%%%%%%%%%%%%%%%%%%%%%%%%%%%%%%%%%%%%%%
%
\end{document}